\documentclass[letterpaper, conference]{ieeeconf}
\IEEEoverridecommandlockouts

\usepackage{lipsum}
\usepackage[sort, compress]{cite}
\usepackage[pdftex]{graphicx}
\usepackage{url}
\usepackage{float}
\usepackage{amsmath,amsfonts,amssymb}
\usepackage{xcolor}
\usepackage{mathtools}
\usepackage{theorem}
\usepackage{algorithm}
\usepackage{algpseudocode}
\usepackage{hyperref}
\usepackage{tabularx}
\usepackage{booktabs}
\usepackage{subcaption}
\usepackage{scalerel}

\newcommand{\nx}{{n_{\mathrm x}}}
\newcommand{\ny}{{n_{\mathrm y}}}
\newcommand{\dnu}{{n_{\mathrm u}}}
\newcommand{\np}{{n_{\mathrm p}}}
\newcommand{\D}{\mathcal{D}}
\newcommand{\Real}{\mathbb{R}}

\newcommand{\x}{\mathrm{x}}
\newcommand{\y}{\mathrm{y}}
\newcommand{\given}{\, | \,}

\DeclareMathOperator{\vectorize}{vec}

{}
{}
{}

\title{\LARGE \bf
Learning Surrogate LPV State-Space Models \\with Uncertainty Quantification*
}
\author{E. Javier Olucha, Amritam Das and Roland T\'{o}th
\thanks{*This research was supported by the European Space Agency (grant number: 4000145530) and The MathWorks Inc. Opinions, findings, conclusions or recommendations expressed in this paper are those of the authors and do not necessarily reflect the views of The MathWorks Inc. or the European Space Agency.}
\thanks{E.J. Olucha, A. Das and R. T\'{o}th are with the Control Systems Group, Eindhoven University of Technology, The Netherlands. R. T\'{o}th is also with the Systems and Control Laboratory, HUN-REN Institute for Computer Science and Control, Hungary. Email addresses: {\{\tt\small{e.j.olucha.delgado, am.das, r.toth\}@tue.nl}}}
\thanks{Corresponding author: E. Javier Olucha}}%

\begin{document}

\maketitle
\thispagestyle{empty}
\pagestyle{empty}

\begin{abstract}
  The \emph{linear parameter-varying} (LPV) framework enables the construction of surrogate models of complex nonlinear and high-dimensional systems, facilitating efficient stability and performance analysis together with controller design. 
Despite significant advances in data-driven LPV modelling, existing approaches do not quantify the uncertainty of the obtained LPV models.
Consequently, assessing model reliability for analysis and control or detecting operation outside the training regime requires extensive validation and user expertise. 
This paper proposes a Bayesian approach for the joint estimation of LPV state-space models, including their scheduling map, together with characterization of the model uncertainty and confidence bounds on the predicted model response directly from input-output data.
Both aleatoric uncertainty due to measurement noise and epistemic uncertainty arising from limited training data and structural bias are considered. 
The resulting model preserves the LPV structure required for controller synthesis while enabling computationally efficient simulation and uncertainty propagation. 
The approach is demonstrated on the surrogate modelling of a two-dimensional nonlinear interconnection of mass-spring-damper systems.
\end{abstract}

\bstctlcite{IEEEexample:BSTcontrol}
\section{Introduction~\label{sec:intro}}
Surrogate models are low-complexity representations that retain essential dynamic information of a system for a given utilization objective. 
In modern engineering workflows, they play a key role in enabling efficient simulation, analysis of system behaviour, and controller design. For example, surrogate models can drastically reduce simulation time when high-fidelity models are computationally expensive~\cite{hou2022}. 
Alternatively, when the available models are overly complex or partially unknown, surrogate representations can be identified from system data, enabling the use of existing tools for analysis or controller design. 

Despite their efficacy, assessing the reliability of surrogate models remains a critical challenge.
In practice, surrogate models are often used under operating conditions different from those observed during training, where their predictions may significantly deviate from the true system response. 
Without explicit \emph{uncertainty quantification} (UQ), extensive validation campaigns with several iterations are required to detect such situations. 
Obtaining uncertainty certificates by accounting for aleatoric and epistemic uncertainties for surrogate models is therefore essential for their safe and reliable deployment in engineering applications. 

Extensive research on UQ for data-driven modelling has been conducted in the system identification, statistics and machine learning communities, with Bayesian formulations becoming prominent in the latter. Bayesian \emph{neural networks} (NNs) and related formulations have been proposed to capture model uncertainty and confidence bounds on the predicted model response~\cite{heckerman2008, kwon2020, kong2020, gawlikowski2023, he2026}. 
More recently, these ideas have been extended to dynamical systems, including nonlinear \emph{state-space} (SS) models~\cite{shapovalova2019, bhusal2019, courts2023, bao2024, dehkordi2025} and neural ordinary differential equations~\cite{zou2024}. 
Despite these advances, such models remain limited in their usefulness for control-oriented applications. In particular, NN models rely on general nonlinear parametrizations that do not provide a structure directly suitable for analysis or controller design with the existing tools. Moreover, the lack of structure prevents the systematic use of local linear approximations for parameter initialisation or the specification of informative priors based on engineering insight.

In contrast, the \emph{linear parameter-varying} (LPV) framework~\cite{toth2010} offers a model structure that can represent nonlinear system behaviour via a scheduling map, while remaining linear for fixed values of the scheduling variables.
In particular, \emph{self-scheduled} LPV models generate the scheduling trajectory based on their own state and input signals, eliminating the need for exogenous scheduling variables.
This enables efficient simulation and the use of existing tools for LPV analysis and controller design~\cite{hoffmann2015, koelewijn2023}; nevertheless, UQ for LPV models remains largely unexplored. 
The only existing approach~\cite{bao2021} characterizes epistemic model uncertainty in LPV--SS identification using Bayesian NNs. Despite its excellent results, it only partly addresses the problem, since it assumes the availability of pre-selected exogenous scheduling variables, which are typically not available in surrogate modelling. Moreover, it employs nonlinear parametrizations of the LPV matrices that do not explicitly enforce the structural properties required for analysis or controller design, and confidence bounds on the model response are not systematically characterized.

The problem addressed in this paper is the identification of self-scheduled LPV--SS surrogate models with UQ from input-output data. 
To this end, we propose a Bayesian approach that enables the joint estimation of the scheduling map, an LPV--SS model affinely dependent on the scheduling variables, and the associated model uncertainty together with confidence bounds on the predicted model response. Both aleatoric uncertainty due to measurement noise and epistemic uncertainty arising from limited training data and structural bias are explicitly accounted for under an output-error noise setting, and the approach directly handles multi-input multi-output systems. 
The proposed formulation estimates models directly suitable for control-oriented applications together with uncertainty certificates, addressing a gap in the current literature. It enables a computationally efficient characterization of the predicted output response in terms of the mean and confidence intervals with complexity that scales linearly with the prediction horizon. Moreover, the LPV structure allows the incorporation of informative parameter priors based on linear approximations.

The paper is structured as follows. In Section~\ref{sec:problem}, the problem of identifying self-scheduled LPV--SS models with UQ from input-output data is formulated. For this problem, the proposed approach for the joint learning of the model parameters and the scheduling together with UQ is presented in Section~\ref{sec:method}. In Section~\ref{sec:examples}, the capabilities of the method are demonstrated on the surrogate modelling problem of a two-dimensional nonlinear mass-spring-damper system. Finally, in Section~\ref{sec:conclusion}, the main conclusions on the achieved results and further research directions are discussed.

\textit{Notation:} The sets of integers and real numbers are denoted as $\mathbb{Z}$ and $\Real$, respectively. The row-wise vectorization of a matrix $M \in \Real^{a \times b}$ is denoted by $\vectorize{(M)} \in \Real^{1 \times ab}$.
The weighted squared 2-norm of a vector $x \in \Real^{n_\mathrm x}$ with weight matrix $W \in \Real^{n_\mathrm x \times n_\mathrm x}$ is defined as $\| x \|_W^2 := x^\top W x$.
\section{Problem definition\label{sec:problem}}
Consider a \emph{discrete-time} (DT) data-generating system defined in terms of the SS representation:
\begin{equation}\label{eq:true_model}
    \mathcal{S} = \left\{ 
    \begin{aligned}
        x_{k+1} &= f(x_k, u_k), \\
        y_k &= h(x_k, u_k) + e_k,
    \end{aligned}
    \right.
\end{equation}
where $k \in \mathbb{Z}$ denotes the DT step, $x_k \in \Real^\nx$ is the state, $u_k \in \Real^{\dnu}$ is the input, $y_k \in \Real^\ny$ is the measured output of the system, and $e_k \in \Real^\ny$ is assumed to be a realization of an i.i.d. Gaussian noise process with zero mean and covariance $\Sigma_{\mathrm e} \in \Real^{\ny \times \ny}$, i.e., $e_k \sim \mathcal{N}(0, \Sigma_{\mathrm e})$. The functions $f:\Real^\nx \times \Real^\dnu \to \Real^\nx$ and $h:\Real^\nx \times \Real^\dnu \to \Real^\ny$ are considered to be real-valued and deterministic. We assume that an \emph{input-output} dataset $\D_N = \{(u_k, y_k)\}_{k=0}^{N}$ is recorded from~\eqref{eq:true_model} for an input sequence $u_{0:N}$, where $u_{0:N} := \{u_k\}_{k=0}^N$, and possibly unknown initial state $x_0 \in \Real^\nx$.

The considered surrogate model for $S$ is sought in an LPV--SS form with \emph{affine} dependency on scheduling variables, given by
\begin{equation}\label{eq:surrogate}
    S_\theta : \left\{
    \begin{aligned}
        \hat{x}_{k+1} &= A(\rho_k, \theta_M) \hat{x}_k + B(\rho_k, \theta_M) u_k, \\
        \hat{y}_k &= C(\rho_k, \theta_M) \hat{x}_k + D(\rho_k, \theta_M) u_k, \\
        \rho_k &= \eta(\hat{x}_k, u_k, \theta_\eta),
    \end{aligned}
    \right.
\end{equation}
where $\hat{x}_k \in \Real^{\hat{n}_{\mathrm{x}}}$ is the state with $\hat{n}_{\mathrm x} < \nx$, $\rho_k \in \Real^{\np}$ is the scheduling variable and $\eta:\Real^{\hat{n}_{\mathrm{x}}} \times \Real^{\dnu} \times \Real^{n_{\theta_\eta}} \to \Real^{\np}$ is the \emph{scheduling map}, parametrized as a \emph{feedforward neural network} (FNN), while the matrix functions $A, \dots, D$, collected as
\begin{equation*}
    M(\rho_k, \theta_M) = \begin{bmatrix}
        A(\rho_k, \theta_M) & B(\rho_k, \theta_M) \\ C(\rho_k, \theta_M) & D(\rho_k, \theta_M),
    \end{bmatrix}
\end{equation*}
depend affinely on $\rho_k$:
\begin{equation}\label{eq:affine}
    M(\rho_k, \theta_M) = M_0 + \sum_{i=1}^{\np}\rho_k^i M_i,
\end{equation}
and the parameters $\theta_M$ are the elements of the matrices in~\eqref{eq:affine}. Therefore, the parameter vector to learn becomes $\theta = \vectorize{([M_0 \ M_1 \cdots M_{\np}], \theta_\eta, \hat{x}_0)} \in \Real^{n_\theta}$, including the initial condition of the surrogate $\hat{x}_0$.
In case prior knowledge of the system is available, the considered LPV--SS model form, including its scheduling map, can alternatively follow model-augmentation structures such as those proposed in~\cite{hoekstra2025}.
The output can be expressed as
\begin{equation*}
    y_k = \hat{y}(k \mid k-1, \theta) + \varepsilon_k,
\end{equation*}
where $\hat{y}(k \mid k-1, \theta)$ is the one-step-ahead predictor associated with~\eqref{eq:surrogate}, which, due to the \emph{output error} (OE) noise setting, is the forward simulated model output, i.e., $\hat{y}(k \mid k-1, \theta) = \hat{y}(k \mid u_{0:k}, \hat{x}_0, \theta)$.
When~\eqref{eq:surrogate} and~\eqref{eq:true_model} correspond to the same dynamics, then $ \varepsilon_k = e_k \sim \mathcal{N}(0, \Sigma_{\mathrm{e}})$, hence the likelihood of a single observation $y_k$ is
\begin{equation}
    p(y_k \given u_{0:k}, \theta, \Sigma_{\mathrm e}) = \mathcal{N} (y_k \given  \hat{y}(k \mid k-1, \theta),\Sigma_{\mathrm{e}}).
\end{equation}
Furthermore, as $\{e_k\}_{k=0}^N$ is i.i.d., the likelihood of $\D_N$ is
\begin{equation}
    p(\D_N \mid \theta, \Sigma_{\mathrm e}) = \prod_{k=0}^N p(y_k \mid u_{0:k}, \theta, \Sigma_{\mathrm e}).
\end{equation}
For $\theta$, we take the prior parameter distribution to be Gaussian:
\begin{equation}
    p(\theta \given \mu_{\mathrm o}, \Sigma_{\mathrm o}) = \mathcal{N}(\theta \given \mu_{\mathrm o}, \Sigma_{\mathrm o}),
\end{equation}
where $\mu_{\mathrm o} \in \Real^{n_\theta}$ and $\Sigma_{\mathrm o} \in \Real^{n_\theta \times n_\theta}$ are the prior mean and covariance matrix for $\theta$, respectively. By Bayes rule, the \emph{parameter posterior distribution} is given by
\begin{equation}\label{eq:posterior_parameter}
    p(\theta \given \D_N, \mu_{\mathrm o}, \Sigma_{\mathrm o}, \Sigma_{\mathrm e}) = \frac{p(\D_N \given \theta, \Sigma_{\mathrm e}) \, p(\theta \given \mu_{\mathrm o}, \Sigma_{\mathrm o})}{p(\D_N \given \mu_{\mathrm o}, \Sigma_{\mathrm o}, \Sigma_{\mathrm e})},
\end{equation}
where the denominator in~\eqref{eq:posterior_parameter} is constant, providing normalization.
The \emph{predictive distribution}, which characterizes the confidence in the predicted model response for a given input sequence $u_{0:k}$, is obtained by marginalizing~\eqref{eq:posterior_parameter}:
\begin{equation}\label{eq:predictive_posterior}
    p(y_k \: | \: u_{0:k}, \D_N) \! = \! \! \int \! \! p(y_k \: | \: u_{0:k}, \theta, \Sigma_\mathrm e) p(\theta \: | \: \D_N, \mu_{\mathrm o}, \Sigma_{\mathrm o}, \Sigma_{\mathrm e}) \mathrm{d}\theta.
\end{equation}
However, this integral is analytically intractable due to the nonlinear dependence of $\hat{y}(k \mid k-1, \theta)$ on $\theta$, and sampling-based approximations are computationally expensive and scale poorly with model dimension, making them unsuitable for the intended surrogate modelling setting.

Given a dataset $\D_N$, our objective is therefore twofold: (i) to estimate $\theta$ of the surrogate~\eqref{eq:surrogate}, and (ii) to compute an efficient approximation of the predictive distribution~\eqref{eq:predictive_posterior}.

In the next section, we develop a computationally efficient approximation of the predictive distribution together with the associated parameter estimation procedure.
\section{Method~\label{sec:method}}
In this section, we address the estimation of the LPV--SS model parameters using a \emph{Maximum a Posteriori} (MAP) formulation. Based on the MAP estimate, a Gaussian approximation of the parameter distribution $p(\theta \given \D_N, \mu_{\mathrm o}, \Sigma_{\mathrm o}, \Sigma_{\mathrm e})$ is derived, which is then used to derive a Gaussian approximation of the predictive distribution $p(y_k \given u_{0:k}, \D_N)$. Finally, computational aspects of the proposed approach are discussed.
\subsection{Maximum a Posteriori point estimate of the parameters\label{sub:MAP}}
To estimate the model parameters we consider the parameter posterior distribution~\eqref{eq:posterior_parameter}. Since the normalization constant is independent of $\theta$,
\begin{equation}\label{eq:posterior_parameter_proportional}
    p(\theta \given \D_N, \mu_{\mathrm o}, \Sigma_{\mathrm o}, \Sigma_{\mathrm e}) \propto p(\D_N \given \theta, \Sigma_{\mathrm e}) \, p(\theta \given \mu_{\mathrm o}, \Sigma_{\mathrm o}).
\end{equation}
Taking the logarithm of~\eqref{eq:posterior_parameter_proportional} yields the objective function
\begin{equation}\label{eq:log_posterior}
    \mathcal{L}(\theta) = \text{cnst} - \frac{1}{2}\sum_{k=0}^{N} \|\varepsilon(k \given \theta) \|_{\Sigma_{\mathrm e}^{-1}}^2 - \frac{1}{2} \|\theta - \mu_{\mathrm o}\|_{\Sigma_{\mathrm o}^{-1}}^2,
\end{equation}
where $\varepsilon(k \given \theta) = y_k - \hat{y}(k \given k-1, \theta)$, and the constant term
\begin{align*}
    &\text{cnst} = -\frac{(N+1)}{2} \ln((2\pi)^{\ny} \det\Sigma_{\mathrm e}) - \frac{1}{2} \ln((2\pi)^{n_\theta} \det \Sigma_{\mathrm o}).
\end{align*}
is independent of $\theta$ and can therefore be omitted in the cost.
Treating $\mu_{\mathrm o}$, $\Sigma_{\mathrm o}$, and $\Sigma_{\mathrm e}$ as fixed, user-defined hyperparameters and maximizing~\eqref{eq:log_posterior} yields the MAP estimate
\begin{equation}\label{eq:MAP_optimization}
    \vectorize{(\theta_{\mathrm{MAP}})} = \arg \max_{\theta} \mathcal{L}(\theta), \vspace{-2pt}
\end{equation}%
subject to 
\begin{equation*}
     \begin{aligned}
        \hat{x}_{k+1} &= A(\rho_k, \theta_M) \hat{x}_k + B(\rho_k, \theta_M) u_k, \\
        \hat{y}_k &= C(\rho_k, \theta_M) \hat{x}_k + D(\rho_k, \theta_M) u_k, \\
        \rho_k &= \eta(\hat{x}_k, u_k, \theta_\eta).
    \end{aligned}
\end{equation*}
Alternatively, these hyperparameters can be estimated by maximizing the marginal likelihood $p(\D_N\given\mu_{\mathrm o},\Sigma_{\mathrm o},\Sigma_{\mathrm e})$.
In the special case of $\Sigma_{\mathrm o}^{-1} \to 0$, the MAP estimate reduces to the classical maximum likelihood estimate.

\subsection{Gaussian approximation of the parameter distribution\label{subsection:gaussian_approx_parameter_posterior}}
The parameter posterior $p(\theta \given \D_N, \mu_{\mathrm o}, \Sigma_{\mathrm o}, \Sigma_{\mathrm e})$ characterizes the model uncertainty associated with the identified parameters.
Due to the nonlinear dependence of $\hat{y}(k \given k-1, \theta)$ on $\theta$, this distribution is non-Gaussian.
To obtain a tractable approximation, we employ the Laplace method and approximate the parameter posterior with
\begin{equation*}
    q(\theta \given \D_N, \mu_{\mathrm o}, \Sigma_{\mathrm o}, \Sigma_{\mathrm e}) = \mathcal{N}(\theta \given \mu_\mathrm{ap}, \Sigma_\mathrm{ap}).
\end{equation*}
To this end, consider a second-order Taylor expansion of the logarithm of~\eqref{eq:posterior_parameter_proportional} around a linearization point $\theta_\ast$:\vspace{-5pt}
\begin{multline}\label{eq:laplace_approx_taylor}
    \ln p(\theta \given \D_N, \mu_{\mathrm o}, \Sigma_{\mathrm o}, \Sigma_{\mathrm e}) \approx \ln p(\theta_\ast \given \D_N, \mu_{\mathrm o}, \Sigma_{\mathrm o}, \Sigma_{\mathrm e}) \\ - Q(\theta - \theta_\ast)  - \frac{1}{2}(\theta - \theta_\ast)^\top P (\theta - \theta_\ast),
\end{multline}
where\vspace{-2pt}
\begin{align*}
    Q &=   -\frac{\partial \ln p(\theta \given \D_N, \mu_{\mathrm o}, \Sigma_{\mathrm o}, \Sigma_{\mathrm e})}{\partial \theta} \big|_{\theta = \theta_\ast}, \quad Q \in \Real^{1 \times n_\theta}, \\
    P &= -\frac{\partial^2 \ln p(\theta \given \D_N, \mu_{\mathrm o}, \Sigma_{\mathrm o}, \Sigma_{\mathrm e})}{\partial \theta \partial \theta^\top} \big|_{\theta = \theta_\ast}, \quad P \in \Real^{n_\theta \times n_\theta}.
\end{align*}
A natural choice is $\theta_\ast = \theta_{\mathrm{MAP}}$, since it maximizes the parameter posterior, implying that the gradient term $Q$ vanishes, and is available from~\eqref{eq:MAP_optimization}.
Exponentiation of~\eqref{eq:laplace_approx_taylor} leads to the unnormalized Gaussian distribution
\begin{equation}\label{eq:unnormal_gauss_param}
     p(\theta) 
     \approx  p(\theta_{\mathrm{MAP}}) \exp(-\frac{1}{2} \|\theta - \theta_{\mathrm{MAP}}\|_{ P}^2),
\end{equation}
where the conditioning on $\D_N, \mu_{\mathrm o}, \Sigma_{\mathrm o}, \Sigma_{\mathrm e}$ is omitted for readability. 
The standard multivariate Gaussian normalization of~\eqref{eq:unnormal_gauss_param} yields\vspace{-5pt}
\begin{equation*}
    p(\theta \given \D_N, \mu_{\mathrm o}, \Sigma_{\mathrm o}, \Sigma_{\mathrm e}) \approx \frac{\det(P)^{1/2}}{(2\pi)^{n_\theta / 2}} \exp (-\frac{1}{2}\|\theta - \theta_{\mathrm{MAP}} \|_{P}^2).
\end{equation*}
The Hessian $P$ can be written as
\begin{equation}\label{eq:hessian}
    P = \Sigma_{\mathrm o}^{-1} + \sum_{k=0}^N J_k^\top \Sigma_{\mathrm e}^{-1} J_k -  \sum_{k=0}^N \sum_{i=1}^\ny \left[\Sigma_{\mathrm e}^{-1} \varepsilon(k \given \theta_{\mathrm{MAP}})\right]_i H_{k,i},
\end{equation}
where
\begin{equation}\label{eq:jacobian}
J_k = \frac{\partial \hat{y}(k \given k-1, \theta)}{\partial\theta}\big|_{\theta = \theta_{\mathrm{MAP}}} \in \Real^{\ny \times n_\theta},
\end{equation}
and $H_{k, i} \in \Real^{n_\theta \times n_\theta}$ denotes the Hessian of the $i$-th component of $\hat{y}(k \given k-1, \theta)$:
\begin{equation*}
    H_{k, i} = \frac{\partial^2 \, \hat{y}(k \given k-1, \theta)_i}{\partial \theta \partial \theta^\top}\big|_{\theta = \theta_{\mathrm{MAP}}}, \quad i=1, \dots, \ny.
\end{equation*}
As in the Gauss--Newton method,~\eqref{eq:hessian} is approximated by neglecting second-order derivative terms, leading to
\begin{equation}\label{eq:hessian_approx}
    P \approx \Sigma_{\mathrm o}^{-1} + \sum_{k=0}^N J_k^\top \Sigma_{\mathrm e}^{-1} J_k.
\end{equation}
This approximation is generally accurate, for instance, when $\varepsilon(k \given \theta)$ is small and uncorrelated with the second-order terms, which is expected at $\theta_\ast = \theta_{\mathrm{MAP}}$. 
Thus, the Gaussian approximation $q(\theta \given \D_N, \mu_{\mathrm o}, \Sigma_{\mathrm o}, \Sigma_{\mathrm e})$ is given by
\begin{equation*}
    \mu_\mathrm{ap} = \theta_{\mathrm{MAP}}, \quad \Sigma_\mathrm{ap} = P^{-1}.
\end{equation*}
\subsection{Gaussian approximation of the predictive distribution}
As discussed above, the predictive distribution~\eqref{eq:predictive_posterior} is analytically intractable and we seek a computationally efficient approximation suitable for surrogate modelling. Following standard approaches in Bayesian learning~\cite[Chapter~5.7]{bishop2006} and recent applications to dynamical systems~\cite{dehkordi2025}, we approximate the predictive distribution also as a Gaussian.

To this end, we combine the Gaussian approximation of the parameter posterior $p(\theta \given \D_N, \mu_{\mathrm o}, \Sigma_{\mathrm o}, \Sigma_{\mathrm e})$ provided in Subsection~\ref{subsection:gaussian_approx_parameter_posterior} with a linearization of the surrogate $S_\theta$. The later is obtained through a first-order Taylor series expansion of $S_\theta$ around $\theta_{\mathrm{MAP}}$:
\begin{multline}
    p(y_k \given u_{0:k}, \theta, \Sigma_{\mathrm e}) \approx \mathcal{N}(y_k \given \hat{y}(k \given k-1, \theta_{\mathrm{MAP}}) \\ 
    + J_k(\theta - \theta_{\mathrm{MAP}}), \Sigma_{\mathrm e}),
\end{multline}
where $J_k$ denotes the Jacobian defined in~\eqref{eq:jacobian}. Using the standard Gaussian marginalization result\footnote{
If $p(x) = \mathcal{N}(x \given \mu, \Sigma_{\mathrm{x}})$ and $p(y \given x) = \mathcal{N}(y \given Ax + b, \Sigma_{\mathrm{c}})$, then
\mbox{$p(y) = \int p(y \given x)\, p(x)\, \mathrm{d}x
    = \mathcal{N}(y \given A\mu + b,\;\Sigma_{\mathrm{c}} + A \Sigma_{\mathrm{x}} A^\top).$}} leads to
\vspace{-5pt}\begin{equation}\label{eq:predictive_posterior_approx}
    \!\! p(y_k | u_{0:k}, \D_N) \approx \! \mathcal{N} \! \left(y_k \given \hat{y}(k \given  k-1, \theta_{\mathrm{MAP}}), \sigma_k^2(u_{0:k})\right),
\end{equation}\vspace{-5pt}
where the variance is given by
\begin{equation}\label{eq:predictive_posterior_var}
    \sigma^2_k(u_{0:k}) = \Sigma_{\mathrm e} + J_k \Sigma_{\mathrm {ap}} J_k^\top,
\end{equation}
and the dependence of $J_k$ on $u_{0:k}$ and $\hat{x}_0$ is implicit through $\hat{y}(k \given k-1, \theta_{\mathrm{MAP}})$, with $\hat{x}_0$ included in $\theta_{\mathrm{MAP}}$.

The approximation~\eqref{eq:predictive_posterior_approx} constitutes the main result of this work: it provides both the predictive mean and its associated uncertainty. The variance in~\eqref{eq:predictive_posterior_var} naturally decomposes into aleatoric $\Sigma_{\mathrm e}$ and epistemic $J_k \Sigma_{\mathrm {ap}} J_k^\top$ uncertainty terms.
Finally, the Gaussian approximation~\eqref{eq:predictive_posterior_approx} enables the construction of confidence regions for the predicted output. 
For $y_k \sim \mathcal{N}(y_k \given \hat{y}_k, \Sigma_k)$, where $\Sigma_k = \sigma^2_k(u_{0:k})$, the $\alpha$-confidence region is given by
\begin{equation*}
    (y_k - \hat{y}_k)^\top \Sigma_k^{-1} (y_k - \hat{y}_k)
    \leq \chi^2_{\ny}(\alpha),
\end{equation*}
where $\alpha \in (0, 1)$ and $\chi^2_{\ny}(\alpha)$ denotes the inverse cumulative distribution function of the chi-squared distribution.
Such confidence regions can, for example, account for model uncertainty in \emph{model predictive control} (MPC), analogous to uncertainty tubes in tube-based MPC.\vspace{-5pt}
\subsection{Computational details}
The computation of the Jacobians $J_k$ can be executed recursively as proposed in~\cite{forgione2023}, resulting in a computational cost that scales linearly in the simulation horizon. Let the state sensitivity be denoted by 
$s_k = \partial \hat{x}_k / \partial \theta$. 
By differentiating~\eqref{eq:surrogate} with respect to $\theta$, the evolution of $s_k$ is described by
\begin{align}\label{eq:Jk}
\frac{\partial \hat{y}_k}{\partial \theta} &= \frac{\partial h_k}{\partial \hat{x}_k} s_k + \frac{\partial h_k}{\partial \theta}, &
    s_{k+1} &= \frac{\partial f_k}{\partial \hat{x}_k} s_k + \frac{\partial f_k}{\partial \theta},
\end{align}
where
\vspace{-5pt}
\begin{align*}
    \frac{\partial h_k}{\partial \theta} &= \! 
    \bigg[\phi_k \frac{\partial}{\partial \theta_M}
    (G_0 \! + \! \sum_{i=1}^{\np} \rho_k^i G_i) \quad \! \! \phi_k \sum_{i=1}^{\np} G_i
    \frac{\partial \rho_k^i}{\partial \theta_\eta} \quad \! \! 0_{\ny \times \hat{n}_{\mathrm{x}}}
    \bigg], \\
    \frac{\partial h_k}{\partial \hat{x}_k}
    &=
    C_0 + \sum_{i=1}^{\np} C_i \rho_k^i
    + \sum_{i=1}^{\np}
    (C_i \hat{x}_k + D_i u_k)
    \frac{\partial \rho_k^i}{\partial \hat{x}_k},
\end{align*}
with $G_i = \vectorize{([C_i, D_i])}$, $\phi_k = [\hat{x}_k^\top ~ u_k^\top] \otimes I_{\ny}$ and initial sensitivity $s_0 = [ 0_{\hat{n}_{\mathrm{x}}\times n_{\theta_{M}}} \quad  0_{\hat{n}_{\mathrm{x}}\times n_{\theta_\eta}}   \quad  I_{\hat{n}_{\mathrm{x}}}]$.
Here, $\otimes$ denotes the Kronecker product, $I_n$ the identity matrix of size $n$, and $0_{n\times m}$ the zero matrix of size $n \times m$.
The partial derivatives of $f_k$ are obtained in a similar manner, and $\partial \rho_k / \partial \theta_\eta$, $\partial \rho_k / \partial \hat{x}_k$ can be computed via backpropagation. 
Then, a single evaluation of~\eqref{eq:Jk} scales asymptotically as\vspace{-7pt}
\begin{multline*}
    J_{\mathrm{cost}} = \mathcal{O}(\np \hat{n}_{\mathrm{x}}^2 + \hat{n}_{\mathrm{x}}^2 n_\theta + \np \hat{n}_{\mathrm{x}} n_{\theta_\eta} + \ny \hat{n}_{\mathrm{x}} n_\theta) \\ + \mathcal{O}(\partial \rho_k / \partial \theta_\eta) + \mathcal{O}(\partial \rho_k / \partial \hat{x}_k).
\end{multline*}
and its recursive evaluation over $T$ steps scales $\mathcal{O}(T J_{\mathrm{cost}})$, which is linear in the time horizon $T$ and in the number of parameters $n_\theta$, and quadratic in the state dimension $\hat{n}_{\mathrm{x}}$.

The Hessian approximation in~\eqref{eq:hessian_approx} can be computed offline as it depends on $\D_N$. Since $\Sigma_{\mathrm{ap}} = P^{-1}$, only its inverse is required in~\eqref{eq:predictive_posterior_var}. However, direct computation of $P^{-1}$ becomes increasingly ill-conditioned as the data grows. 
Instead, $P^{-1}$ can be computed recursively using numerically stable updates based on the Woodbury matrix identity~\cite{higham2002}. Specifically, for $k = 1, \dots, N$ with $P_0 = \Sigma_{\mathrm{o}}^{-1}$:
\vspace{-3pt}\begin{equation}\label{eq:woodbury_hessian_inv_approx}
    P_{k+1}^{-1} = P_k^{-1} - P_k^{-1} J_k^\top(\Sigma_\mathrm e + J_k P_k^{-1} J_k^\top)^{-1} J_k P_k^{-1}.
\end{equation}
Each update incorporates the information from one data point in $\D_N$, yielding $\Sigma_{\mathrm{ap}} = P_{N+1}^{-1}$.
\section{Simulation results~\label{sec:examples}}
In this section, the effectiveness of the proposed method is demonstrated on an identification problem of a surrogate model for a two-dimensional \emph{mass-spring-damper} (MSD) system. The considered MSD system can represent mechanical trusses, cantilevers or discrete approximations of continuous media. The objective is to learn a reduced-order surrogate LPV model together with UQ from measured input-output data, rather than to capture the complete underlying system dynamics.
The implementation used to produce these results is available at \mbox{\href{https://gitlab.com/Javi-Olucha/lpv-sysid-uq}{https://gitlab.com/Javi-Olucha/lpv-sysid-uq}}.
\subsection{System description}
\begin{figure}[t]
    \centering
    \includegraphics[width=0.6\columnwidth]{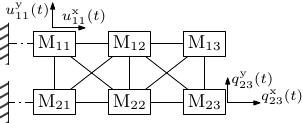}
    \caption{Graphic representation of the two-dimensional interconnection of mass-spring-damper systems.\vspace{-19pt}}
    \label{fig:benchmark}
\end{figure}
The two-dimensional MSD system described in~\cite{olucha2026} is considered.
As depicted in Fig.~\ref{fig:benchmark}, the system consists of six point masses of $0.5~\mathrm{kg}$ each arranged in a $2 \times 3$ rectangular grid.
For each mass $(i,j)$,  with $i \in \{1, 2\}$ and $j \in \{1, 2, 3\}$, the state vector $( q^\x_{ij} \ q^\y_{ij} \ \dot{q}^\x_{ij} \ \dot{q}^\y_{ij} )^\top$ describes its position and velocity in the horizontal and vertical directions, accounting for a total of 24 states.
Each mass is connected to its adjacent neighbours via springs and dampers. The masses in the first column are additionally connected to an infinitely rigid wall.
The connections with the cardinal neighbours and the wall act exclusively along the horizontal and vertical directions. The diagonal connections depend on the Euclidean distance between the corresponding masses.

The spring and damper force relations considered in this example are defined as follows.
For the wall connections,
\vspace{-3pt}\begin{equation*}
    k^\x_{\mathrm{wall}} = q^\x_{i1}, \ k^\y_{\mathrm{wall}} = 2q^\y_{i1}, \ d^\x_{\mathrm{wall}} = \dot{q}^\x_{i1}, \ d^\y_{\mathrm{wall}} = \dot{q}^\y_{i1}.
\end{equation*}
For the cardinal connections,
\vspace{-3pt}\begin{gather*}
    k^\x_{\mathrm{card}} = q^\x_{ij,ls}, \ k^\y_{\mathrm{card}} = q^\y_{ij,ls} + (q^\y_{ij,ls})^3, \\
    d^\x_{\mathrm{card}} = \dot{q}^\x_{ij,ls}, \ d^\y_{\mathrm{card}} = \dot{q}^\y_{ij,ls},
\end{gather*}
where $q_{ij,ls} = q_{ls} - q_{ij}$. 
For the diagonal connections,
\vspace{-3pt}\begin{equation*}
    k_{\mathrm{diag}} = 5 \tanh{(r_{ij, ls})}, \quad d_{\mathrm{diag}} = 0.5 \sin{(\dot{r}_{ij, ls})}, 
\end{equation*}
where $r_{ij,ls} = \|q_{ij, ls}\|_2$.
Lastly, an external input force $u(t) = ( u^\x_{11} \ u^\y_{11})^\top$ is applied to the first mass, while the position of the last mass, i.e., $w(t) = ( q^\x_{23} \ q^\y_{23})^\top$, is considered to be the system output. The reader is referred to~\cite{olucha2026} for a detailed description of the equations of motion.

The system dynamics are discretized using a fourth-order Runge--Kutta (RK4) method with sampling time $T_{\mathrm s} = 0.05~\mathrm s$ and a zeroth-order hold on the input. 
The resulting DT representation of the system in the form of~\eqref{eq:true_model} is given by
\begin{equation}\label{eq:DT_benchmark}
    \begin{aligned}
        {x}_{k+1} &= f(x_k, u_k), \quad
        w_k = h(x_k, u_k),
    \end{aligned}
\end{equation}
where $x_k\in \Real^{24}$ is the state, $u_k \in \Real^2$ is the input and $w_k \in \Real^2$ is the noise-free output.
\subsection{Experiment design}
\begin{figure}[t]
\centering
\begin{subfigure}{\columnwidth}
\includegraphics[width=\columnwidth, trim={0 1 0 1}, clip]{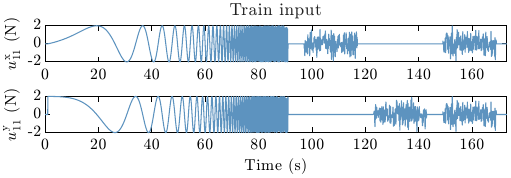}
    \label{fig:train_input}
\end{subfigure}
\begin{subfigure}{\columnwidth}
\vspace{-10pt}
\includegraphics[width=\columnwidth, trim={0 3 0 3}, clip]{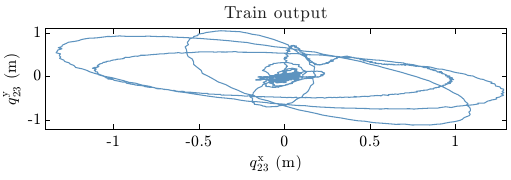}
    \label{fig:train_output}
\end{subfigure}
\vspace{-30pt}
\caption{Input and output trajectories of the training data set.\vspace{-15pt}}
\label{fig:train_data}
\end{figure}
Two data sets are generated from~\eqref{eq:DT_benchmark}, denoted by $\D_{\mathrm{train}}$ and $\D_{\mathrm{test}}$, used for training and testing the surrogate models, respectively. For $\D_{\mathrm{train}}$ the DT model is simulated with
\begin{equation}\label{eq:train_u}
    u_{\mathrm{train}}(k) =     
    \begin{bmatrix}
        2\varphi_{1, \mathrm{x}}(k) + \varphi_{2, \mathrm{x}}(k) \\
        2\varphi_{1, \mathrm{y}}(k) + \varphi_{2, \mathrm{y}}(k)
    \end{bmatrix},
\end{equation}
from zero initial conditions, resulting in 3460 data points ($172.95~\mathrm{s}$).
The components $\varphi_{1, \bullet}$ are chirp input forces $\varphi_{1, \bullet}(k) = \sin{((1.4\pi)^{T_\mathrm s k/90} T_\mathrm s k + \phi_{\bullet})}$, active for $0 \leq k T_{\mathrm s} \leq 90$, where $\phi_{\mathrm{x}} = 0$ and $\phi_{\mathrm{y}} = \pi/2$. 
The components $\varphi_{2, \bullet}$ are multi-sine random phase input forces $\varphi_{2, \bullet}(k) = \sum_{j=1}^{n_{\mathrm r}} 0.1\sin{(2 \pi \omega_j T_{\mathrm s} \; k + \phi_j)}$ providing separate and simultaneous excitation of the input channels. Specifically, $\varphi_{2, \mathrm x}$ is active over the time intervals $k T_\mathrm s \in [96, 116] \cup [148, 168]$, while $\varphi_{2, \mathrm y}$ is active over $k T_\mathrm s \in [122, 142] \cup [148, 168]$.
The frequencies $\{\omega_j\}_{j=1}^{n_{\mathrm r}}$ lie in $[0, 1.4 \pi)$ with resolution of $0.05~\mathrm{Hz}$ and the phases satisfy $\phi_j \sim \mathcal{U}(0, 2\pi)$. 
The measured output is obtained by adding white noise to the noise-free output $w_k$, i.e., $y_k = w_k + e_k$, where $e_k \sim\mathcal{N}(0, \Sigma_{\mathrm e})$ with covariance $\Sigma_{\mathrm e} = \operatorname{diag}(0.3795, 0.262)$, corresponding to a \emph{signal-to-noise} (SNR) ratio of $35~\textrm{dB}$. 
The resulting input and output trajectories of $\D_{\mathrm{train}}$ are shown in Fig.~\ref{fig:train_data}. The train dataset $\D_{\mathrm{train}}$ is then scaled and normalized. 

For $\D_{\mathrm{test}}$, the DT model is simulated with
\begin{equation}
    u_{\mathrm{test}}(k) = \begin{bmatrix}
        \varphi_{1, \x}(k) + \varphi_{2, \x}(k) \\
        \varphi_{1, \y}(k)
    \end{bmatrix},
\end{equation}
from zero initial conditions, resulting in 600 data points ($29.95~\mathrm s$). 
The input components are defined as $\varphi_{1, \x}(k) = 2 \sin{(1.6 \pi T_{\mathrm s} k + \pi/3)}$, 
$\varphi_{2,\x}(k)$ is a step of magnitude $2$ over the interval $k T_{\mathrm s} \in [1,4]$,
and $\varphi_{1, \y}(k) = \sin{(2 \pi 0.1 T_{\mathrm s} k + \pi/2)} - 2 \sin{(2 \pi 0.05 T_{\mathrm s} k + \pi / 7)}$. 
The measured output is defined as $y_k = w_k + e_k$, where $e_k \sim \mathcal{N}(0, \Sigma_{\mathrm e})$ is included with covariance $\Sigma_e = \operatorname{diag}(0.1483, 0.1919)$, corresponding to a SNR ratio of $35~\mathrm{dB}$. 
Under these settings, $\D_{\mathrm{test}}$ also challenges the extrapolation capability of the learned surrogate. In particular, $\varphi_{1, \x}$ contains higher-frequency components than those present in $u_{\mathrm{train}}$, while $\varphi_{2, \x}$ induces a step response, exciting dynamics of different nature. 
\begin{figure}[t]
\centering
\begin{subfigure}{\columnwidth}
\includegraphics[width=\columnwidth, trim={0 0 0 0}, clip]{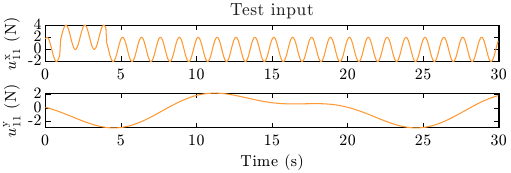}
    \label{fig:val_input}
\end{subfigure} 
\begin{subfigure}{\columnwidth}
\vspace{-10pt}
\includegraphics[width=\columnwidth, trim={0 2 0 3}, clip]{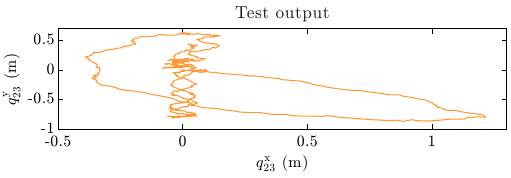}
    \label{fig:val_output}
\end{subfigure}
\vspace{-30pt}
\caption{Input and output trajectories of the test data set.\vspace{-12pt}}
\label{fig:val_data}
\end{figure}
Lastly, as shown in Fig.~\ref{fig:val_data}, the resulting output trajectories of $\D_{\mathrm{test}}$ explore further on the lower region of the output space.
\vspace{-2pt}
\subsection{Selection of informative parameter priors}\label{sec:init_prior}
The selection of parameter prior distributions is a critical and non-trivial aspect of Bayesian estimation.
For NN-based model structures, incorporating engineering insights into informative priors is generally difficult, limiting the effective use of the Bayesian framework. 
The considered LPV model structure partially alleviates this issue, as priors for the LTI part of the LPV model, characterized by $M_0$, can be obtained systematically via linear system identification.
In particular, a \emph{best linear approximation}~\cite{SCHOUKENS2020310} (BLA) can provide an informative choice for the prior mean of $M_0$, while the associated covariance is chosen heuristically to reflect the uncertainty in this estimate.

To this end, three linear DT LTI--SS models are identified using $\D_{\mathrm{train}}$.
The \textsc{Matlab} functions \textsc{ssest} and \textsc{n4sid} are used to estimate the models $S_{\textrm{ssest}}$ and $S_{\textrm{n4sid}}$, respectively, using the \emph{prediction error minimization} (PEM) and subspace methods~\cite{ljung1999}. A third model $S_{\textrm{jpem}}$ is estimated using the \textsc{Jax}-based~\cite{bradbury2018} PEM approach~\cite{bemporad2025}. 
In all cases, the sampling time is set to $T_{\mathrm s} = 0.05~\mathrm{s}$, the state dimension to $\hat{n}_x = 6$, the feedthrough matrix fixed to zero, the initial state $\hat{x}_0$ is estimated, and the noise model estimation is disabled.

The obtained models are evaluated in simulation using their respective estimated initial conditions, both for the train input $u_{\mathrm{train}}$ and the test input $u_{\mathrm{test}}$. As a measure
\begin{table}[ht]
\centering
\caption{BFR of the estimated LTI models}\vspace{-5pt}
\label{tab:BFR_LTI}
\begin{tabularx}{0.95\columnwidth}{XXXX}
BFR (\%)                 & $S_{\textrm{ssest}}$ & $S_{\textrm{n4sid}}$ & $S_{\textrm{jpem}}$ \\ \midrule
$u_{\mathrm{train}}$ & 37.40                & 25.28                & 85.38 \\
$u_{\mathrm{test}}$  & -265.62              & -175.39              & 67.84 \\ \bottomrule
\end{tabularx}\vspace{-5pt}
\end{table}
\begin{figure}[H]
    \centering
    \vstretch{0.9}{\includegraphics[width=0.85\columnwidth, trim={15 0.5 0 3}, clip]{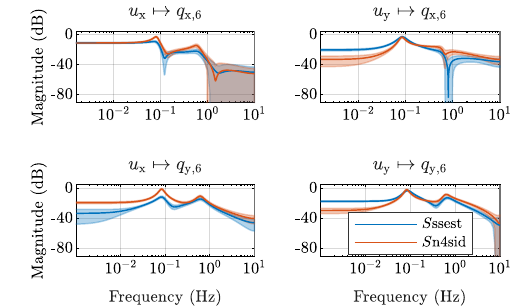}}
    \vspace{-3pt}
    \caption{Frequency response of the estimated LTI models $S_{\mathrm{ssest}}$ and $S_{\mathrm{n4sid}}$ where the shaded area shows the uncertainty interval for confidence level 95\%.\vspace{-12pt}}
    \label{fig:LTI_freqresp_confidence}
\end{figure}
\begin{figure}[H]
    \centering
    \includegraphics[width=\columnwidth, trim={2 4 2 3.5}, clip]{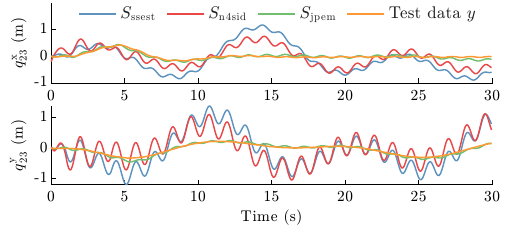}
    \vspace{-17pt}
    \caption{Simulated output trajectories of the estimated LTI models for $u_{\mathrm{test}}$.\vspace{-5pt}}
    \label{fig:init_prior_LTI}
\end{figure}
of model quality, the simulation \emph{best fit rate} (BFR)\footnote{\small $\mathrm{BFR} = \left(1 - \sqrt{\frac{\sum_{k=0}^{N} \|y(k) - \hat{y}(k) \|_2^2}{\sum_{k=0}^{N} \|y(k) - y_{\mathrm{mean}} \|_2^2}}\right)\cdot 100\%$, where $y$ is the data sequence, $y_{\mathrm{mean}}$ is the sample mean of $y$, and $\hat{y}$ is the predicted response of the model.} is reported in Tab.~\ref{tab:BFR_LTI}. 
Additionally, the confidence region corresponding to one standard deviation in the frequency response of $S_{\textrm{ssest}}$ and $S_{\textrm{n4sid}}$ is shown in~\ref{fig:LTI_freqresp_confidence}, and the simulation results for $u_{\mathrm{test}}$ are displayed in Fig.~\ref{fig:init_prior_LTI}.
The results show that $S_{\mathrm{jpem}}$ significantly outperforms the other models for $u_{\mathrm{train}}$ and is the only model that accurately approximates the true output for $u_{\mathrm{test}}$. Moreover, the confidence regions of $S_{\textrm{ssest}}$ and $S_{\textrm{n4sid}}$, computed from the parameter covariance of the identified LTI models, fail to reflect the performance degradation on the test data, highlighting the limitations of classical LTI uncertainty estimates and motivating the use of surrogate models with UQ.
\subsection{Learning an LPV State-Space model with UQ}
Now, an LPV-SS model with UQ, denoted by $S_{\mathrm{lpv}}$, is identified using the proposed approach, which is implemented in \textsc{Python} with the \textsc{Jax} library. We define a discrete-time LPV model structure in the form of~\eqref{eq:surrogate} with sampling time $T_{\mathrm s} = 0.05~\text{s}$ and a state, input, output and scheduling variable dimensions of $\hat{n}_\mathrm x = 6$, $\dnu = 2$, $\ny = 2$ and $n_{\mathrm p} = 1$, respectively. The feedthrough matrix is constrained to zero, and the scheduling map $\eta(\hat{x}_k, u_k)$ is parametrized by a FNN with two hidden layers and three fully connected neurons per layer, using $\tanh$ activation functions.
The resulting LPV model contains a total of $n_\theta = 168$ to-be-trained parameters.

The hyperparameters are taken as follows. The prior mean $\mu_{M_0}$ associated with the LTI part is obtained from the BLA model $S_{\mathrm{jpem}}$ identified in Section~\ref{sec:init_prior}, with covariance $\Sigma_{\mathrm{o}, M_0} = 0.25 I$, reflecting the high confidence indicated in Tab.~\ref{tab:BFR_LTI}.
The prior mean and covariance for the remaining parameters are set to zero and $10 I$, respectively, reflecting the absence of prior structural knowledge while providing mild regularization toward the origin.
The measurement noise covariance is set to $\Sigma_{\mathrm e} = 100 I$. The model parameters are initialized accordingly: the elements of $M_0$ are set to the values of $S_{\mathrm{jpem}}$, the entries of the scheduling-dependent matrices $M_i$ are drawn from a zero-mean normal distribution, and the FNN weights are initialized using the Xavier method~\cite{glorot2010}.

The model parameters are then estimated as detailed in Section~\ref{sub:MAP} by solving~\eqref{eq:MAP_optimization} using the strategy in \cite{bemporad2025}, with $2000$ ADAM~\cite{kingma2017a} iterations followed by up to $6000$ L-BFGS~\cite{byrd1995a} iterations. The optimization is repeated $16$ times from different random initial guesses for all model parameters except $M_0$, which is initialized from the BLA model, resulting in a total training time\footnote{On a laptop with an i7-13850HX (2.10 GHz) CPU and 64 GB RAM.} of $\approx 51~\text{s}$.
Next, the parameter posterior is approximated via the Laplace method from Section~\ref{subsection:gaussian_approx_parameter_posterior}, and the covariance $\Sigma_{\mathrm{ap}}$ is computed using~\eqref{eq:woodbury_hessian_inv_approx}, requiring $\approx 1~\text{s}$.
Using the resulting model $S_{\mathrm{lpv\text{-}uq}}$ and covariance $\Sigma_{\mathrm{ap}}$, the predictive distribution in~\eqref{eq:predictive_posterior_approx} is evaluated to compute the model response $\hat{y}(k \mid k-1, \theta_{\mathrm{MAP}})$ and associated variance $\sigma^2_k$.
For visualization, the diagonal entries of $\sigma^2_k$ are used to construct $\pm 2\sigma_{k,i}$ bounds around the corresponding predicted output at each time step.
\begin{table}[t]
\centering
\caption{BFR and computation time of the simulated model response together with the $2\sigma$ uncertainty bounds.}
\label{tab:BFR_LPV_UQ}
\begin{tabularx}{0.99\columnwidth}{XXX}
Dataset & BFR (\%)               & Computation time (s) \\ \midrule
$u_{\mathrm{train}}$ & 96.46 & 0.287\\ 
$u_{\mathrm{test}}$ & 86.96 & 0.254\\\bottomrule
\end{tabularx}
\end{table}
\begin{figure}[t]
    \centering
\includegraphics[width=\columnwidth,  trim={4 0 0 0}, clip]{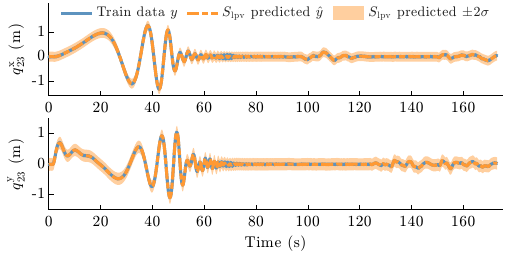}
    \caption{Simulation of the estimated $S_{\mathrm{lpv}}$ for $u_{\mathrm{train}}$, showing the predicted mean and the $\pm 2 \sigma$ confidence interval of the model response together with the measured train output.\vspace{-15pt}}
    \label{fig:lpv_uq_train}
\end{figure}

\begin{figure}[t]
    \centering
\includegraphics[width=\columnwidth, trim={2 0 0 0}, clip]{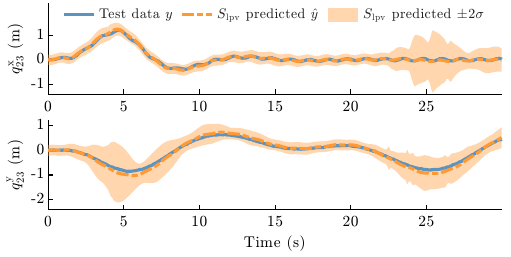}\vspace{-5pt}
    \caption{Simulation of the estimated $S_{\mathrm{lpv}}$ for $u_{\mathrm{test}}$, showing the predicted mean and $\pm 2 \sigma$ confidence interval of the model response together with the measured test output.\vspace{-16pt}}
    \label{fig:lpv_uq_validation}
\end{figure}

The simulation performance of the surrogate is evaluated for both $u_{\mathrm{train}}$ and $u_{\mathrm{test}}$, and the computation time and simulation BFR are reported in Tab.~\ref{tab:BFR_LPV_UQ}. 
The results indicate that the identified LPV model achieves superior simulation performance compared to linear counterparts, while maintaining a computational cost that scales linearly with the prediction horizon. The simulation results with confidence intervals corresponding to two standard deviations for $u_{\mathrm{train}}$ and $u_{\mathrm{test}}$ are displayed in Fig.~\ref{fig:lpv_uq_train} and Fig.~\ref{fig:lpv_uq_validation}, respectively. The predicted confidence intervals increase at time instances with larger prediction error, indicating the estimated uncertainty appropriately reflects the model reliability.
\vspace{-5pt}

\section{Conclusion\label{sec:conclusion}}
This paper introduced a novel Bayesian approach for learning LPV-SS surrogate models that jointly estimate the scheduling map, the system dynamics, and the model uncertainty. 
Both aleatoric uncertainty, arising from measurement noise, and epistemic uncertainty due to limited training data and structural bias are considered, enabling the computation of confidence bounds on the predicted model response. 
The experimental results confirm that, as is well-known is system identification, uncertainty estimates obtained from linear identification approaches are unreliable when the underlying data-generating system
can not be fully represented by the model structure. In contrast, the proposed method provides meaningful uncertainty characterization while achieving superior prediction performance.
For future research, we plan to investigate more accurate approximation of the posterior predictive distribution to further improve the accuracy of the characterization of confidence bound on the model response.
\vspace{-5pt}
\vspace{-12pt}
\section*{Acknowledgements}
\vspace{-3pt}
The authors thank Valentin Preda (European Space Agency) for contributing to the original idea of this work.
\vspace{-5pt}

\bibliographystyle{IEEEtran}
\bibliography{learning-LPV-SS-with-UQ}

\addtolength{\textheight}{-12cm}   


\end{document}